\documentclass[manuscript]{aastex}

\slugcomment{}

\shorttitle{}
\shortauthors{Umana et al.}

\begin{document}

\title{EVLA observations of the nebula around G79.29+0.46}


\author{G. Umana, C. S. Buemi, C. Trigilio and P. Leto }
\affil{INAF-Osservatorio Astrofisico di Catania, Via S. Sofia 78, 95123 Catania, Italy
}
\email{Grazia.Umana@oact.inaf.it}
\and
\author{C. Agliozzo  and A. Ingallinera}
\affil{Universit\'a di Catania  and INAF-Osservatorio Astrofisico di Catania, Via S. Sofia 78, 95123 Catania, Italy
}
\author{A. Noriega-Crespo}
\affil{Spitzer Science Center, California Institute of Technology, Mail Code 314-6, Pasadena, CA 91125, USA}
\and
\author{J. L.  Hora}
\affil{Harvard-Smithsonian Center for Astrophysics, 60 Garden St. MS-65, Cambridge, MA 02138-1516, USA}

%\affil{Harvard Smithsonian Center for Astrophysics, 60 Garden St. MS-65, Cambridge, MA02138-1516}


%\altaffiltext{1}{Visiting Astronomer, Cerro Tololo Inter-American Observatory.
%CTIO is operated by AURA, Inc.\ under contract to the National Science
%\altaffiltext{2}{Society of Fellows, Harvard University.}
%\altaffiltext{3}{present address: Center for Astrophysics,
%    60 Garden Street, Cambridge, MA 02138}
%\altaffiltext{4}{Visiting Programmer, Space Telescope Science Institute}
%\altaffiltext{5}{Patron, Alonso's Bar and Grill}


\begin{abstract}

We have observed the radio nebula surrounding the Galactic LBV candidate
G79.29+0.46 with the EVLA at 6~cm.
These new radio observations allow a morphological comparison between 
the radio emission, which traces the ionized gas component, and the mid-IR  emission, a tracer of the 
dust component.  The IRAC ($8 \, \mu m$) and MIPS ($24\, \mu m$ and $70\, \mu m$)  images have been reprocessed and compared with the EVLA map.
We confirm the presence of a second shell at 24$ \mu m$ and also provide evidence for its detection at $70\, \mu m$.
The differences between the spatial morphology of the radio and mid-IR maps 
 indicate the existence of two dust populations, the cooler one emitting mostly at longer wavelengths.
 Analysis of the two dusty, nested shells have provided us with an estimate of the characteristic timescales
 for shell ejection, providing  important constraints for stellar evolutionary models.
   
Finer details of the ionized gas distribution can be 
appreciated thanks to the improved quality of the new 6~cm image,  most notably the highly structured texture of the
nebula.  Evidence of interaction between the nebula and the surrounding interstellar medium 
can be seen in the radio map, including  brighter features 
that delineate  regions where the shell structure is locally modified.
In particular, the brighter filaments in the south-west region appear to frame the shocked southwestern clump reported from 
CO observations. 

\end{abstract}

\keywords{circumstellar matter--- infrared: stars --- stars: early-type ---stars: individual (G79.29+0.46) ---stars: winds, outflows}



\section{Introduction}
Massive stars play a fundamental role in the evolution of galaxies.
They are major contributors to the interstellar UV radiation and, via their strong stellar winds,
provide enrichment of processed material (gas and dust) and mechanical energy to the interstellar medium.
Despite their importance, the details of post-MS evolution of massive stars are still poorly understood.
Recent evolutionary models suggest that Luminous Blue Variables (LBVs)  and related 
transition objects may play a key role in the massive star evolution, representing
a crucial phase during which a star loses most of its H envelope  \citep{lamers01}. 
 More recently, it has also been pointed out that LBVs might be direct Supernovae progenitors \citep{Vink_06, Smith_08},
enhancing their importance in the framework of stellar evolution

LBVs are luminous (intrinsically bright, $ L \sim 10^{6} \,{\rm L_{\odot}}$) objects,
exhibiting different kinds of photometric and spectroscopic variabilities. They are massive
($M \sim 20-120 \,{\rm M_{\odot}}$), characterized by  intense mass-loss rates ($10^{-6}-10^{-4}\,{\rm M_{\odot} yr^{-1}}$), which can also occur in the form of eruptive events. 
Although eruptive events have been witnessed very rarely (e.g., $\eta$ Car and P Cyg) the presence of extended, dusty  circumstellar nebulae around LBVs  (LBVNe) suggests that they are a common aspect of LBV behavior \citep{Weis08}.
There are, however, many aspects of LBV phenomenon that are not completely understood. Among these are the total mass lost during the LBV phase (a key parameter 
necessary to test evolutionary models), the origin and shaping of the LBVNe, and how  the mass-loss behavior (single versus multiple events, bursts) is related to the physical parameters of the central object.

The mass-loss archeology  of the central object can be recovered  from an analysis of its associated nebula.
A successful approach is based on a synergistic use of different techniques, at different wavelengths, that allows one to analyze the several emitting components coexisting in the nebula. 
In particular, a detailed comparison of mid-IR and radio  maps, with comparable spatial resolution, has provided estimates of both ionized gas and dust masses and allowed us
to sort out morphological differences in the maps which can be associated 
with mass-loss behaviour during the LBV phase \citep{Buemi_2010, Umana_2010}.

\subsection{The nebula surrounding G79.29+0.46}
\noindent

G79.29+0.46 is  considered a LBV candidate because its observed properties to date do not meet the requirements of spectral and photometric variability to be accepted as a bona-fide member.
 However, $H_{\alpha}$ variations, suggestive of mass-loss variability during S-Dor variations, have recently been  reported by \cite{Vink_08}.

The highly symmetric ringlike structure surrounding G79.29+0.46 was first pointed out
by \citet{Wendker_91}. The thermal nature of the continuum radio emission was determined by \cite{Higgs_94},
who  suggested that the ringlike nebula is an ionized shell of swept-up interstellar
material. By examining the IRAS high-resolution images, \citet{Waters_96} concluded that the ringlike nebula is a detached shell formed during an epoch of high mass loss ($ \sim 5 \times 10^{-4}\,{\rm M_{\odot} yr^{-1}}$) followed by a quieter period. This scenario is consistent with G79.29+0.46 being a LBV, where different mass loss events may have occurred in the recent past.

More recently, the Spitzer Space Telescope has provided high-sensitivity, high-resolution IRAC and MIPS maps \citep[The Cygnus-X Spitzer legacy program,][]{Hora_2010} which provides
a better understanding of the dust properties \citep{Kraemer_2010}. Moreover, the detection 
of CO millimetre emission in G79.29+0.46 \citep{Rizzo_08} demonstrates that another component, consisting of molecular gas, is present in the surroundings of this star and should be included in the budget of the total mass lost from the central object. 
In their CO maps of G79.29+0.46, \cite{Rizzo_08} identified components of warm and dense molecular gas whose morphology closely resembles that of the extended IRAS  nebula. They also reported the presence of a shock front, interpreted as a natural consequence of different wind regimes during the central object's evolution.

Despite the wealth of new mid-IR and mm observations of the sources, 
our knowledge of the radio emission has been limited to
the 1988 VLA data reported by \citet{Higgs_94} 
and on the 1400 and 350 MHz Westerbork observations, carried out between 1996 and 1997 and reported by \citet{Setia_03}.
In this paper, we present new EVLA observations with sufficient dynamical range, sensitivity, and angular resolution to provide a good match to the Spitzer images. The EVLA observations allow, for the first time, a detailed morphological comparison with other maps tracing the different components of the nebula.

%\section{The nebula associated to HD~168625}
%%%%%%%%%%%%%%%%%%%%%%%%%%%%%%%%%%%%%%%%%%%%%%%%%%%%%%%%%%%%%%%%%%%%%%%%%%%%%%%%%%%%%%%%%%%%%%%%%%%%%%%%%%%%%%%%%%%%%%%%%%%%%%%%%%%%%%%%%%%
%%%%%%%%%%%%%%%%%%%%%%%%%%%%%%               OSSERVAZIONI E RIDUZIONE                   %%%%%%%%%%%%%%%%%%%%%%%%%%%%%%%%%%%%%%%%%%%%%%%%%%%
%%%%%%%%%%%%%%%%%%%%%%%%%%%%%%%%%%%%%%%%%%%%%%%%%%%%%%%%%%%%%%%%%%%%%%%%%%%%%%%%%%%%%%%%%%%%%%%%%%%%%%%%%%%%%%%%%%%%%%%%%%%%%%%%%%%%%%%%%%%
\section{Observations and Data Reduction}
\subsection{EVLA observations}
Radio observations of G79.29+0.46 and its associated nebula were carried out with the EVLA\footnote{The National Radio Astronomy Observatory is a facility of the National Science Foundation
operated under cooperative agreement by Associated Universities, Inc. } on 2010 June 1 and 11  and 2010 December 1 and 5.
The source was observed at two frequencies (1.4 and 5 GHz), with a total bandpass of 256 MHz and in two different array configurations (D and C).
For each configuration and frequency, the same observing scheme was followed; namely, a 5 minute on-source scan preceded and followed by a 1 minute scan on the phase calibrator,
for a total of 75 minutes on-source integration time.
The source J0137+3309 was used to fix the absolute amplitude scale and to correct for the bandpass. Table 1 contains a summary of the observing details.
The data reduction was carried out within the {\bf C}ommon {\bf A}stronomical {\bf S}oftware {\bf A}pplications (CASA) package, version 3.0.2. At each frequency, the data  from each configuration were first independently calibrated and then combined 
into a single UV data set. The imaging process was performed by setting the Briggs  robust parameter equal to 0,  a compromise between uniform weighting of the baseline for highest angular resolution and natural weighting for highest sensitivity. We also used 
a multi-scale CLEANing algorithm, intended for high resolution image but sensitive to extended structures \citep{Brogan_06}, resulting in a single image with a rms noise of  0.07 ${\rm mJy \,beam^{-1}}$ and a synthetic beam of  4\farcs56$ \times $3\farcs09 for the 5~GHz observations, and a rms noise of  0.5 ${\rm mJy \, beam^{-1}}$ and a synthetic beam of 18\farcs5$ \times$ 18\farcs6 for the 1.4 GHz observations.

\subsection{Spitzer data}
\subsubsection{IRAC data}
Infrared imaging of the field including G79.29+0.46 was performed at 3.6, 
4.5, 5.8, and 8.0~$\mu$m with the InfraRed Array Camera (IRAC) \citep{fazio04} on the Spitzer 
Space Telescope \citep{werner04}.  All available data for G79.29+0.46 
from the cryogenic Spitzer mission archive 
were used, including AORIDs 6050560, 17330688, 27106560, and 27107584.  The observations used the 12 and
30 second HDR modes, which obtain integrations with frame times of 0.6 \& 12 seconds and 
1.2 \& 30 seconds, respectively.
The Basic Calibrated Data (BCD) were retrieved from the archive 
(pipeline version S18.18), and additional interactive processing was performed using the 
{\tt imclean}\footnote{http://irsa.ipac.caltech.edu/data/SPITZER/docs/dataanalysistools/tools/\\\
contributed/irac/imclean/}
tools to remove image artifacts from bright sources, including muxbleed, banding,
and column pulldown effects \citep{hora04}. 
Mosaics at each wavelength were constructed using 
IRACproc \citep{schuster06}, combining the data which was taken at different spacecraft
orientations and resampling to a final image pixel size of 0\farcs6. Outliers due to 
cosmic rays or instrument artifacts not previously flagged are removed in this process. 
The images were combined using a weighting based on the integration time of the frames. 
In pixels that are saturated in the longer exposures, only the shorter frames are used in the final
image.

\subsubsection{MIPS  data}

The MIPS data was taken as part of the Cygnus-X Spitzer Legacy program (PID 40184, PI J. Hora) in the fast scanning
mode with a cross scan step of $148^{\prime\prime}$ to fill the $70\,\mu m$ array, resulting in an integration time of 15.7 seconds per pixel 
on a single scan leg at both 24 and 70$\, \mu m$. The $24\, \mu m$  data were reprocessed using the MIPSGAL \citep{Carey09} 
data processing pipeline \citep{Mizuno08}. The 70$\, \mu m$ data also used the MIPSGAL pipeline \citep{Paladini11} with the 
exception that the non-linearity correction at 70$\, \mu m$  was done per pixel using the prescription defined by 
the SINGS Legacy team \citep{Dale07},  based on the behavior of the MIPS 70$\, \mu m$ calibrators.

\section{The radio nebula associated with G79.29+0.46}
As we intend to compare the spatial distribution of the ionized gas component, traced by the radio observations, with the morphology
of the dust component, traced by the mid-IR observations, in the following we will analyze and discuss only the 5 GHz dataset, as
that image provides details with a spatial resolution comparable to that of Spitzer observations.
The morphology of the radio nebula is evident in our 6~cm EVLA multi-configuration image,
 which reveals a well defined shell-like structure whose overall shape was previously reported by \cite{Higgs_94}.
 However, the improved quality of our image, shown in Figure~\ref{fig_EVLA}, allows us to discern finer
details of the ionized gas distribution, most notably the highly structured texture of the
nebula.  There is evidence for interaction between G79.29+0.46 and the surrounding interstellar medium: i.e., the bright frontal structures in the northeast and the southwest regions of the nebula
that delineate regions where the shell structure is locally modified.
In particular, the brighter filaments in the south-west region appears to frame the shocked southwestern clump observed in CO \citep{Rizzo_08, Jimenez_2010}.

The central object is well-detected, as is the (probably extragalactic)
background object to the 
northwest.
The position, flux density and angular sizes of these components have been 
derived  by fitting two dimensional Gaussian
brightness distributions to the map. We obtained a flux density of $1.51 \pm 0.08$ and $1.88 \pm 0.09$ mJy   for the central and background components, respectively.
The uncertainty associated with the flux density estimation is given by:
\begin{equation}
\sigma=\sqrt{(RMS_{\mathrm{tot}})^{2}+(\sigma_{\mathrm{cal}}S_{\nu})^{2}} 
\end{equation} 
where $RMS_{\mathrm{tot}}$ is the rms noise in the map, and $\sigma_{\mathrm{cal}}$ is the systematic error due to the flux calibrator 
(typically on the order of 3$\%$). 

Within the errors,  the derived flux densities are in agreement with those determined by \cite{Higgs_94},  who derived a spectral index of $1.39 \pm 0.14$ between 5 and 8.4 GHz.
They pointed out that, even though this spectral index is steeper than the 
canonical $0.6$, it is still consistent with mass-loss from a stellar object.

Assuming that the central radio source is related to the stellar wind from the LBV, we can
derive its current-day mass-loss from the standard formula  {\citep{Panagia_75}:
\begin{equation}
\dot {M}= 6.7 \times 10^{-4} v_\infty
F_{\nu}^{3/4} D_\mathrm{kpc}^{3/2} (\nu\times g_\mathrm{ff})^{-0.5}
~~~~ \mathrm M_{\sun}\mathrm{yr}^{-1}
\end{equation}
where full ionization and cosmic abundances have been assumed, $F_{\nu}$ is the observed radio flux density, in  mJy, and $v_{\infty}$ is the terminal velocity of the wind in  $\rm{km\,s}^{-1} $.
The  free-free  Gaunt factor $g_\mathrm{ff}$  is  approximated by
$ g_\mathrm{ff}=9.77(1+0.13 \log{\frac{T^{3/2}}{\nu}})$  \citep{Leitherer_91}.\\
From the radio flux density observed at 5 GHz,  assuming as   stellar wind velocity a value of  $v\sim 110~  \mathrm{km~s}^{-1}$  \citep{Voors_00}, a wind
temperature of $10^4$K and a distance of  1.7 kpc \citep{Jimenez_2010}, we
derive  a mass loss rate of
$\dot {M}= 5 \times 10^{-7}\,\mathrm M_{\sun}\mathrm{yr}^{-1}$.\\
 \cite{Vink_08} pointed out the possibility that  G79.29+0.46 is  associated with the nearby DR15 region rather than with  Cyg OB2. If this is the case,  G79.29+0.46 is likely to be closer, at about $1 \mathrm{kpc}$.
Since mass-loss scales with the distance as $D^{3/2}$, this would imply a reduced mass-loss of a factor 0.45. 
These values are  smaller, but still consistent, with previous evaluations obtained by other authors using
different techniques \citep[e.g.,][]{Waters_96}, but it is at least an order of magnitude smaller than the current-day mass-loss rates derived from radio measurements for
other LBVs \citep{Umana_2010}.

\section{ Ionized component versus dust components}
%\subsection{ By superimposition of different maps }
The $8 \,\mu$m IRAC and  $24$   and $70\, \mu m$ MIPS  images of G79.29+0.46 are shown in Figure \ref{fig_EVLA_MIPS}a.
Previous versions of these images were presented by \cite{Kraemer_2010} and \cite{ Jimenez_2010}, who both pointed out the presence of a second larger radius shell in the $24\,\mu$m image.
Our reprocessed maps confirm the existence of the second $24\,\mu$m shell and, thanks to our improved reduction,  also provide  a hint of its presence at $70\,\mu$m.
Both the $24$ and $70\,\mu$m maps show the same overall distribution of the dust, consistent with at least two nested dusty shells surrounding the LBV.
On the contrary, in the $8\,\mu$m map where a major contribution from warm dust is expected, the dust appears to be more concentrated in two south-west and south-east regions.
 The new EVLA radio observations allow, for the first time, a morphological comparison between the ionized gas and the dust.
In Figure \ref{fig_EVLA_MIPS}b, the same maps are shown but now with the 6~cm EVLA map superimposed using white contours.
Once again there is a difference between the IRAC and MIPS maps:  while the nebular emission at both $24\,\mu$m and $70\,\mu$m is more extended than the ionized gas (radio nebula), the $8\,\mu$m emission appears well-contained 
within the ionized part of the nebula.
Moreover, while the radio emission from the nebula is incomplete in the north region, the emission at $24\,\mu$m and $70\,\mu$m has a more uniform distribution.
The apparent lack of dust in the southern regions in both $8\,\mu$m and $24\,\mu$m are perhaps due to an absorption effect caused by the infrared dark cloud (IRDC) which may be located in front of the the southern region of the nebula.

To better visualize the spatial distribution of  ionized gas relative to the dust,  surface brightness profiles
have been extracted along cuts through the nebula from the 6~cm, $24\,\mu$m and $70\,\mu$m maps.
However, a direct comparison can  be performed only between the 6~cm and the $24\,\mu$m maps, as they share a comparable angular resolution, $ \sim 5^{\prime \prime}$
for the 6~cm EVLA versus $ \sim 6^{\prime \prime}$ for the $24\,\mu$m MIPS
map.  A superposition of the 6~cm EVLA map on the $24\,\mu$m map is shown in Figure~\ref{fig_EVLA_MIPS_24}.
We extracted eighteen cuts from each map and determined an azimuthally-averaged source profile, shown in  (Figure~\ref{profiles}).
The radio shell has a smaller thickness and a sharp decreasing trend in the outer part of the nebula. The dust nebula is more extended and shows a smoother distribution in
 the outer regions. The second shell is clearly evident in the $24\,\mu$m profile, at a distance of about $ \pm 200^{\prime \prime}$ from the center. At the same distance,
 less defined peaks are visible in the $70\,\mu$m profile, smoothed out by the coarser spatial resolution.

\section{Discussion and Conclusions}
In this letter we present new radio EVLA observations of the nebula surrounding the LBV G79.29+0.46.
In particular,  we have analyzed the 6~cm map and, for the first time, compared the radio free-free, which traces the spatial distribution of the ionized gas,
with maps recently obtained in the mid-IR, which traces the spatial distribution of the dust component.

The dust distribution observed at $8\,\mu$m appears different than the distribution at $24$ and $70\,\mu$m. The warmer dust, traced by the $8\,\mu$m
emission, is more concentrated in the southwest and southeast regions.  Moreover, when compared with the radio nebula, the $24$ and $70\,\mu$m emission
appears more extended, while the $8\,\mu$m emission is well-contained inside the radio nebula. 
This is consistent with the existence of two dust components: a cooler component traced by the $24$ and $70\,\mu$m emission, and a warmer dust component, probably consisting of smaller grains, traced by the $8\,\mu$m emission. This warmer component, with morphological properties quite different from  those observed at longer wavelengths, could be related to the PAH emission that has been claimed by \citet{Jimenez_2010}.

An improved MIPS $70\,\mu$m map, constructed using the most updated pipeline from MIPSGAL team,  detects 
the presence of a second shell very similar to that seen at $24\,\mu$m.
Nebular lines and dust continuum are the main possible contributors to the $24\,\mu$m emission from objects
embedded in a dusty nebula with a hot central component.  The possibility of such a combination, 
with one kind of emission being predominant, has been suggested by \citet{Mizuno10} to explain the $24\,\mu$m emission detected in the MIPSGAL Bubbles.
In the case of G79.29+0.46, the presence of similar shells at $24\,\mu$m and at $70\,\mu$m, together with the lack
of any prominent emission line falling within the response curve of the MIPS $24\,\mu$m band,  as evident from the low-resolution IRS spectrum
of the shell \citep{Jimenez_2010}, leaves very little doubt that the shell emission is entirely due to thermal dust emission.

The fact that there are at least two nested shells provides strong constraints on the the origin of the nebula, as it is 
difficult to explain the presence of multiple shells in the hypothesis of a nebula consisting of swept-up ISM material, as suggested by
\cite{Higgs_94}. It is evident that the dusty shells consist of material ejected by the central object in different mass-loss episodes. 

From the $24\,\mu$m emission profile, we have determined that the inner shell peaks at $100^{\prime \prime}$ and the second one at $200^{\prime \prime}$ from the central object. Such dust emission peaks can be related to the epoch when enhanced mass-loss took place.
Assuming a distance of 1.7 kpc \citep{Jimenez_2010}, this corresponds to a linear distance of $0.82$ pc and  $1.64$ pc, respectively.
This implies, assuming a shell expansion velocity of $\sim 30\,\rm{km\,sec}^{-1}$ \citep{Waters_96}, that the two mass-loss episodes occurred  $2.7 \times 10^{4}$  and $5.4 \times 10^{4}$
years ago,  or $1.6  \times 10^{4}$  and $3.1  \times 10^{4}$ years ago, if a distance of 1 kpc, as suggested by \cite{Vink_08}, is used.  
Our results point out that mass-loss can occur in different episodes, for which we derived characteristics timescales, and pose  quite strong constraints that stellar evolution models must take into account.

The direct comparison of the 6~cm and $24\,\mu$m maps (that share comparable spatial resolution) indicates that only part
of the inner, brighter nebula is ionized,  (i.e. the nebula is ionization bounded).  The radio nebula shows a sharper profile, depicting regions where probable interaction
between the second and the first dusty shells is taking place. This is evident in several regions, most notably in the northeast and in the southwest part of the nebula,
where the overall quite regular shell-like morphology is locally disturbed. The mid-IR dust emission depends linearly on density and as a modified blackbody on temperature; the mid-IR emission is therefore most 
sensitive to dust temperature. On the other hand, assuming an uniform temperature, 
the thermal radio emission depends on the square of the ionized gas density. This different dependence on density
means that the radio emission will emphasize features that have the largest density.
This is consistent with the spatial coincidence of the bright radio features with the CO emission reported by \citet{Rizzo_08} and \citet{Jimenez_2010}. In particular, the high critical density of the $J=3-2$ CO line ($ n_{3} \sim 10^{4} \,\rm{cm}^{-3}$) indicates that the CO emission traces higher density regions, confirming that the brighter radio features delineate interactions between the shells where shocks can occur.
\acknowledgments
This work is based in part on observations made with the Spitzer Space Telescope, which is operated by the Jet Propulsion Laboratory, California Institute of Technology under a contract with NASA. Support for this work was provided by NASA through an award issued by JPL/Caltech.

\clearpage
%%%%%%%%%%%%%%%%%%%%%%
%%%%%%%%TABLE%%%%%%%%%%%
\begin{table}
\begin{center}
\caption{Observational summary}
\begin{tabular}{cccccc}

\hline\hline\\
   Observing  & EVLA &  Frequency & Integration time & Phase Cal \\
   dates           & Conf. &   GHz            & (min)                    &                    \\
  \hline\\
                       &            &                        &                             &                     \\
 2010 June 1  &   D      &  1.4                 &            90            & J2052+3635\\
  2010 June 11  &   D      &  4.9                &            90            & J2015+3710\\
  2010 Dec. 1  &   C    &  1.4                 &            90            & J2015+3710 \\
  2010 Dec.  5  &   C      &  4.9              &            90            & J2015+3710\\
     \hline\hline\\
\end{tabular}
\end{center}
%%%%%%%%%%%%%%End Of the TABLE%%%%%%%%%%%%%%%%%%%%%%%%%%%
\end{table}

\clearpage

\begin{figure}
  \includegraphics[width=20cm]{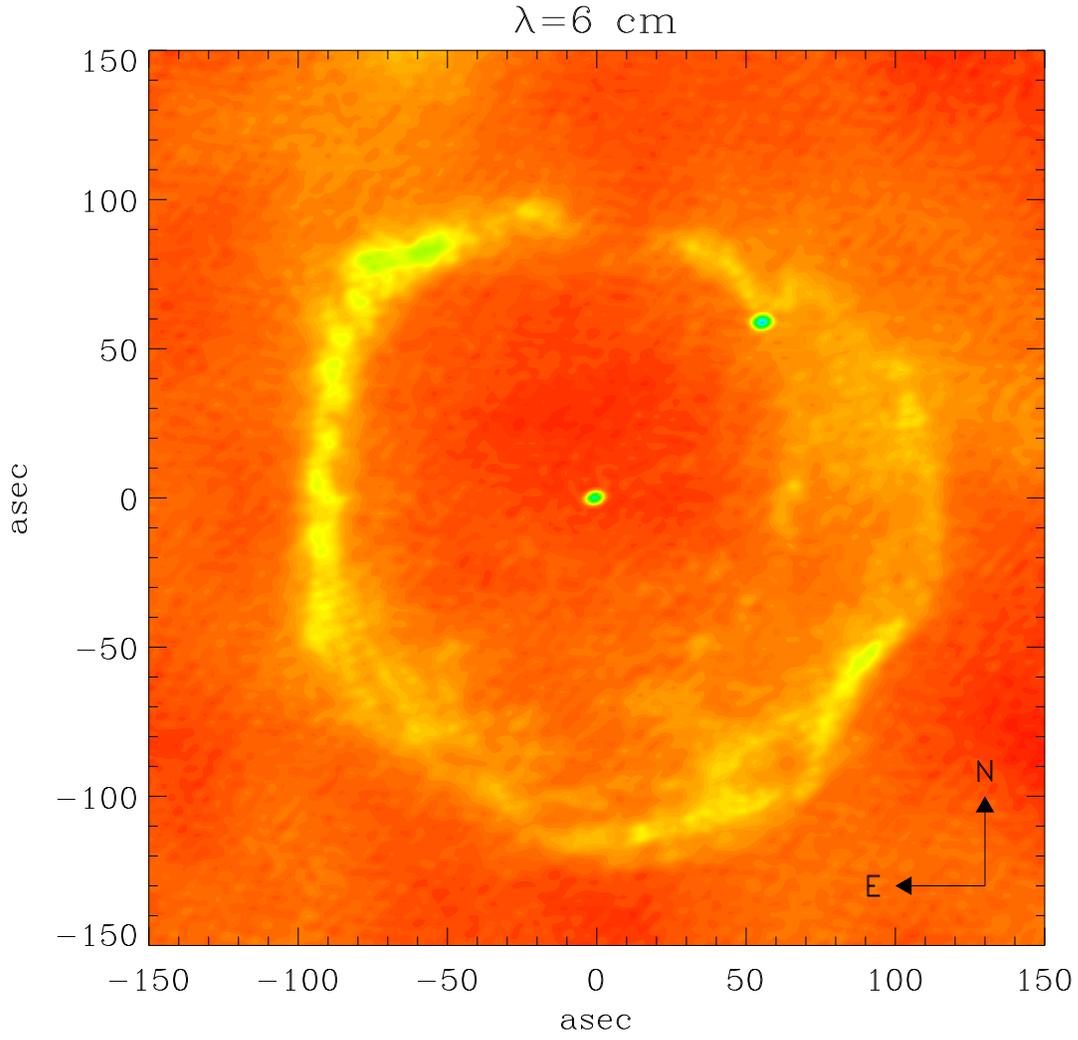}
  % \plotone{g79_c.eps}
\caption{The 6~cm EVLA map of G79.29+0.46 obtained by combining the data from two different array configurations, C and D.
The field of view is $2.5^{\prime} \times 2.5^{\prime}$ centered on the LBV position. }
\label{fig_EVLA}       % Give a unique label
\end{figure}

\clearpage

\begin{figure}
  \includegraphics[width=18cm]{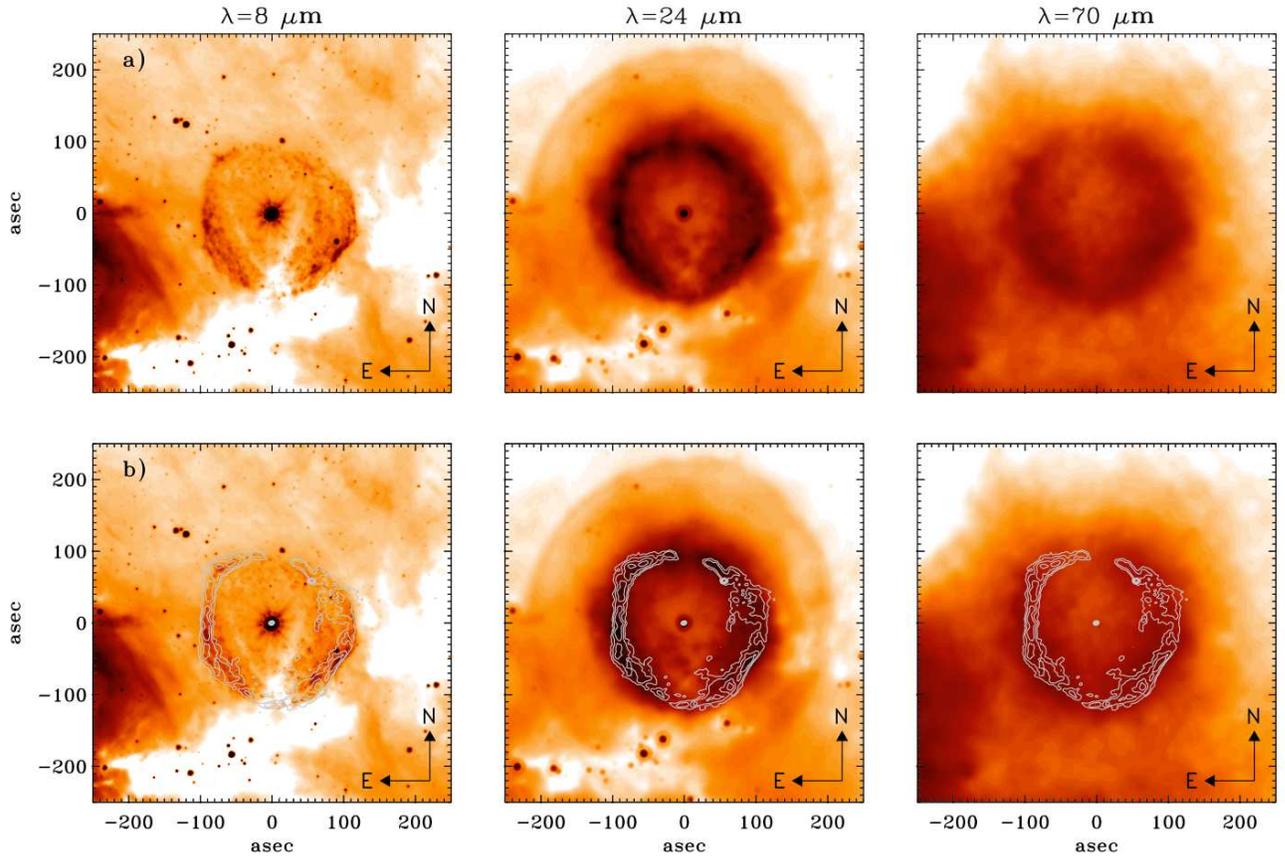}
\caption{Panel a: The $8\,\mu$m  IRAC (right),  the $24\,\mu$m  (center) and the $70\,\mu$m (left) MIPS maps of G79.29+0.46.
Panel b): the same as panel a, but the EVLA 6~cm map   (white contours) superimposed.
To better point out the major radio features, only contours starting from $ 10\%$ of flux density of  the central radio sources have been drawn.  
For  each map, the field of view is $4.2^{\prime} \times 4.2^{\prime}$ centered on the LBV position. 
}
\label{fig_EVLA_MIPS}       % Give a unique label
\end{figure}
\clearpage

\clearpage

\begin{figure}
  \includegraphics[width=16cm]{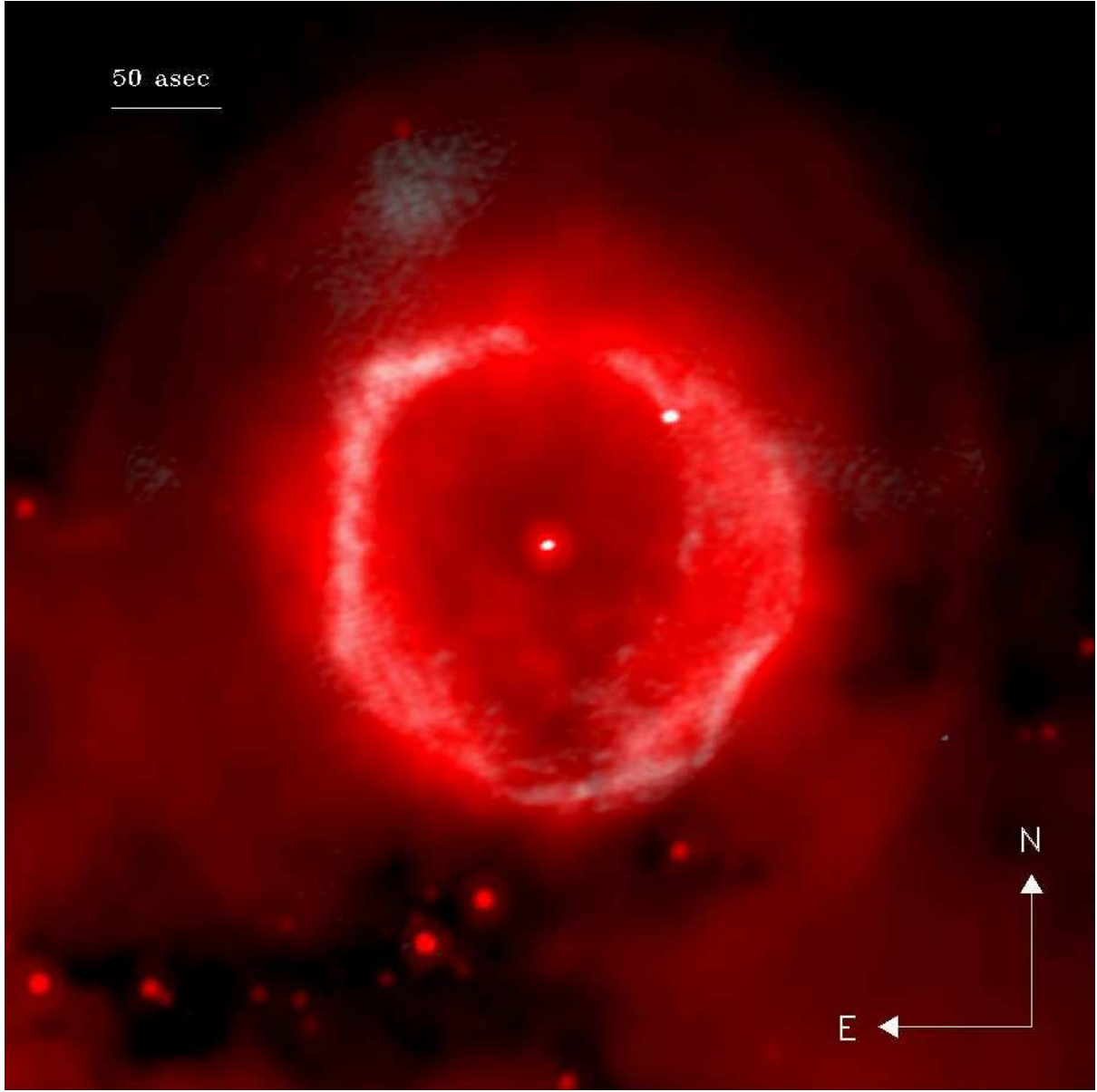}
\caption{The 6~cm EVLA map of G79.29+0.46 (grey) superimposed to the MIPS  $24\,\mu$m map (red).
The field of view is $3.5^{\prime} \times 3.5^{\prime}$ centered on the LBV position.  
}
\label{fig_EVLA_MIPS_24}       % Give a unique label
\end{figure}
\clearpage
\begin{figure}
\includegraphics[width=14cm]{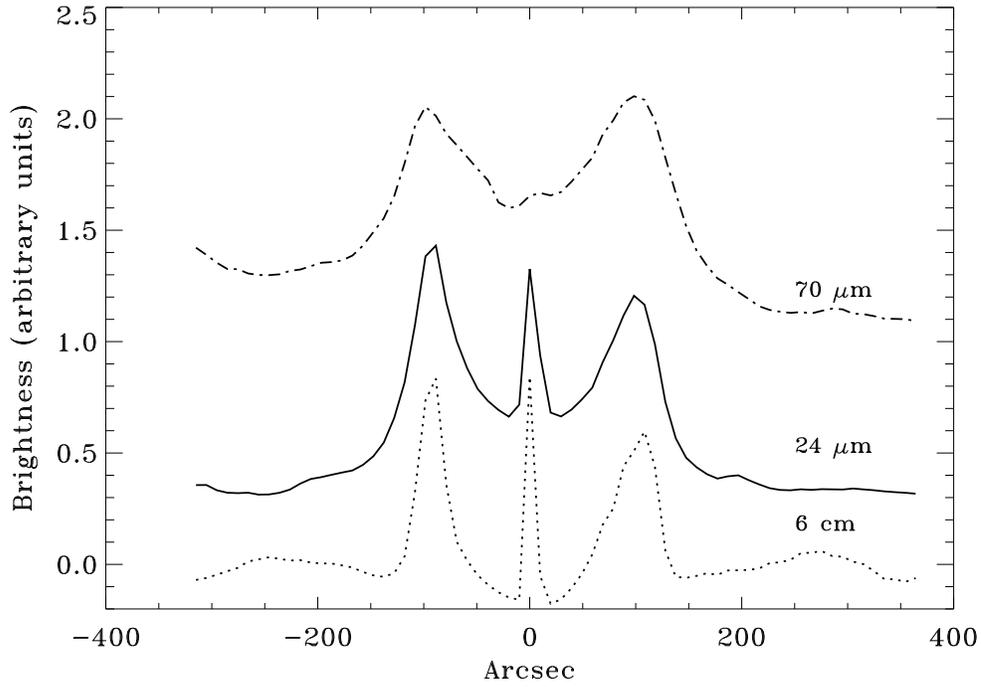}
\caption{Averaged profiles through the nebula at 6~cm (dotted line), $24\,\mu$m (continuous line) and $70\,\mu$m
(dotted slash line).
Each profile has been obtained from 18 individual cuts in the corresponding map
and has been shifted vertically by an arbitrary quantity for easier comparison.}
\label{profiles}       % Give a unique label
\end{figure}

\end{document}